\documentclass{article}

\usepackage{arxiv}

\usepackage[utf8]{inputenc} 
\usepackage[T1]{fontenc}    
\usepackage{hyperref}       
\usepackage{url}            
\usepackage{booktabs}       
\usepackage{amsfonts}       
\usepackage{nicefrac}       
\usepackage{microtype}      
\usepackage{lipsum}		
\usepackage{graphicx}
\usepackage[round]{natbib}
\usepackage{doi}

\usepackage{gensymb}

\usepackage{array,multirow,makecell}
\usepackage{fancybox}
\usepackage{colortbl}
\usepackage{float}

\usepackage{tikz}
\usepackage{multicol}
\usepackage{multirow}
\usepackage{booktabs}

\usepackage[ruled, norelsize, linesnumbered]{algorithm2e}
\usepackage{algpseudocode}

\usepackage{balance}

\usepackage{amsmath} 
\usepackage{amssymb}  

\usepackage{subfig}

\usepackage[percent]{overpic}
\usepackage{xcolor}
\usepackage{rotating}
\usepackage{cases}

\renewcommand{\dblfloatsep}{0pt} 
\renewcommand{\dbltextfloatsep}{0pt} 
\renewcommand{\intextsep}{0pt} 

\allowdisplaybreaks 

      \def\theenumi{\roman{enumi}}

\def\aujour{\number\day \space \ifcase\month\or
janvier\or f�vrier\or mars\or avril\or mai\or
juin\or juillet\or ao�t\or septembre\or octobre\or
novembre\or d�cembre\fi \space \number\year}

\def\cH{{\cal H}}

\def\cL{{\cal L}}

\newtheorem{remark}{Remark}
\newtheorem{ass}{Assumption}
\newtheorem{lemma}{Lemma}

\def\C{{\setbox0=\hbox{$\displaystyle{\rm C}$}
        \hbox{\hbox to0pt{\kern 0.4\wd0\vrule height 0.95\ht0\hss}\box0}}}
\def\Q{{\setbox0=\hbox{$\displaystyle{\rm Q}$}%
    \hbox{\raise 0.2\ht0\hbox to0pt{\kern 0.4\wd0\vrule height
    0.85\ht0\hss}\box0}}} 
\def\R{\mathop{\rm I\mkern -3.5mu R}} 

\def\Rm{{\R}^m}

\def\Rn{{\R}^n}

\def\Rp{{\R}^p}

\def\Rq{{\R}^q}

\def\cH2{{\cal H}_2} 

\def\cL2{\mathop{\mathcal L}_{2}} 

\def\cRH2{\mathop{\cal R \cal H}_2} 
\def\cRL2{\mathop{\cal R \cal L}_{2}} 

\DeclareMathOperator*{\diag}{diag}

\newcommand{\norm}[1]{\left\|{#1}\right\|}
\newcommand{\abs}[1]{\left|{#1}\right|}

\makeatletter
\DeclareRobustCommand\sfrac[1]{\@ifnextchar/{\@sfrac{#1}}
                                            {\@sfrac{#1}/}}
\def\@sfrac#1/#2{\leavevmode\kern.1em\raise.5ex
         \hbox{$\m@th\fontsize\sf@size\z@
                           \selectfont#1$}\kern-.1em
         /\kern-.15em\lower.25ex
          \hbox{$\m@th\fontsize\sf@size\z@
                            \selectfont#2$}}
\makeatother

\usepackage{epstopdf}
\usepackage{rotating}
\usepackage{units}
\usepackage{siunitx}
\usepackage{float}

\title{Data-Driven State Estimation for Light-Emitting Diode Underwater Optical Communication}

\author{ 
\href{}{\includegraphics[scale=0.06]{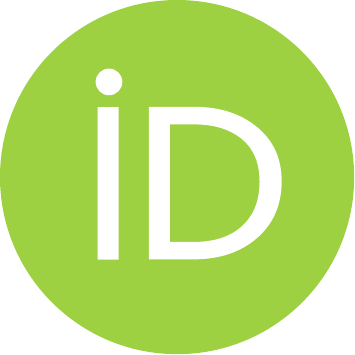}\hspace{1mm}Y.~Li$^1$}, 
\href{}{\includegraphics[scale=0.06]{orcid.pdf}\hspace{1mm}Z.~Liang$^1$}, 
\href{https://orcid.org/0000-0002-2576-5515}{\includegraphics[scale=0.06]{orcid.pdf}\hspace{1mm}I.~N'Doye$^1$}, 
\href{}{\includegraphics[scale=0.06]{orcid.pdf}\hspace{1mm}X. Zhang$^2$}, 
\href{https://orcid.org/0000-0003-4827-1793}{\includegraphics[scale=0.06]{orcid.pdf}\hspace{1mm}M.-S. Alouini$^1$}, 
\href{https://orcid.org/0000-0001-5944-0121}{\includegraphics[scale=0.06]{orcid.pdf}\hspace{1mm}T.-M.~Laleg-Kirati$^1$}
\thanks{This work has been supported by the King Abdullah University of Science and Technology (KAUST) through Base Research Fund (BAS/1/1627-01-01).} \\
$^1$Computer, Electrical and Mathematical Sciences and Engineering Division (CEMSE)\\
King Abdullah University of Science and Technology (KAUST)\\
Thuwal 23955-6900, Saudi Arabia \\
$^2$KAUST and University of Notre Dame, Notre Dame, IN 46556 USA\\
	\texttt{yingquan.li@kaust.edu.sa; zhenwen.liang@kaust.edu.sa;ibrahima.ndoye@kaust.edu.sa} \\
	\texttt{xiangliang.zhang@kaust.edu.sa;slim.alouini@kaust.edu.sa; taousmeriem.laleg@kaust.edu.sa} \\
}

\renewcommand{\shorttitle}{\textit{arXiv} Template}

\hypersetup{
pdftitle={A template for the arxiv style},
pdfsubject={q-bio.NC, q-bio.QM},
pdfauthor={David S.~Hippocampus, Elias D.~Striatum},
pdfkeywords={First keyword, Second keyword, More},
}

\begin{document}
\maketitle

\begin{abstract}
Light-Emitting Diodes (LEDs) based underwater optical wireless communications (UOWCs), a technology with low latency and high data rates, have attracted significant importance for underwater robots. However, maintaining a controlled line of sight link between transmitter and receiver is challenging due to the constant movement of the underlying optical platform caused by the dynamic uncertainties of the LED model and vibration effects. Additionally, the alignment angle required for tracking is not directly measured and has to be estimated. Besides, the light scattering propagates beam pulse in water temporally, resulting in time-varying underwater optical links with interference.  We address the state estimation problem by designing an LED communication system that provides the angular position and velocity information to overcome the challenges. In this way, we leverage the power of deep learning-based observer design to explore the LED communication's state space properly. Simulation results are presented to illustrate the performance of the data-driven LED state estimation.
\end{abstract}

\keywords{Light-emitting diode (LED) \and Underwater optical wireless communication (UOWC) \and Online estimation \and Observer design \and Nonlinear systems \and Deep learning algorithm \and Neural networks. }

\section{Introduction}
Underwater optical wireless communication (UOWC) has rapidly developed nowadays along with the increasing demand for communication such that underwater research scientific, underwater sensors, submarine interconnection, and remote access vehicles. Firstly, a low attenuation window for a blue-green optical beam ensures the stability of the communication link \citep{schirripa2020underwater}. Light beam with a wavelength around 532nm can pass almost 100 percent of the way through water several meters below the surface \citep{gkoura2014underwater}. Thus, constructing a high speed and stable underwater communication link tens of meters long is possible \citep{cossu2019recent}. For instance, an LED-based UOWC system for autonomous underwater vehicles (AUVs) network that reaches over $30$ meters length has recently been designed by \citep{tian2013design}.  Secondly, optical communication is possible to achieve Gb/s or even Tb/s level bandwidth~\citep{wang2018cost},\citep{arvanitakis2020gb}. Thirdly, UOWC saves energy, so it is easy to expand to a large scale. The transmitter only needs a few watts or tens of watts to achieve stable transmission of huge data streams\citep{son2018study}. Hence, the cost to construct an underwater interconnection network or Internet of Things (IoT) is low.

In the coming 6G era, high speed, low power consumption, and long-distance underwater communication scheme are the key factors enhancing the underwater communication \citep{sticklus2018optical}. Traditional acoustic communication generates signals of tens or even hundreds of watts \citep{jiang2018overview}. However, acoustic systems consume massive power and produce severe noise. Additionally,  the transmission rate is only up to tens of kbps \citep{kumara2021underwater}. On the other hand, radio frequency (RF) communication cannot be used to construct long underwater links due to attenuation over $150$ dB/m \citep{saini2017path}.

\begin{figure}[!t]
        \centering
        \begin{overpic}[scale=0.20]{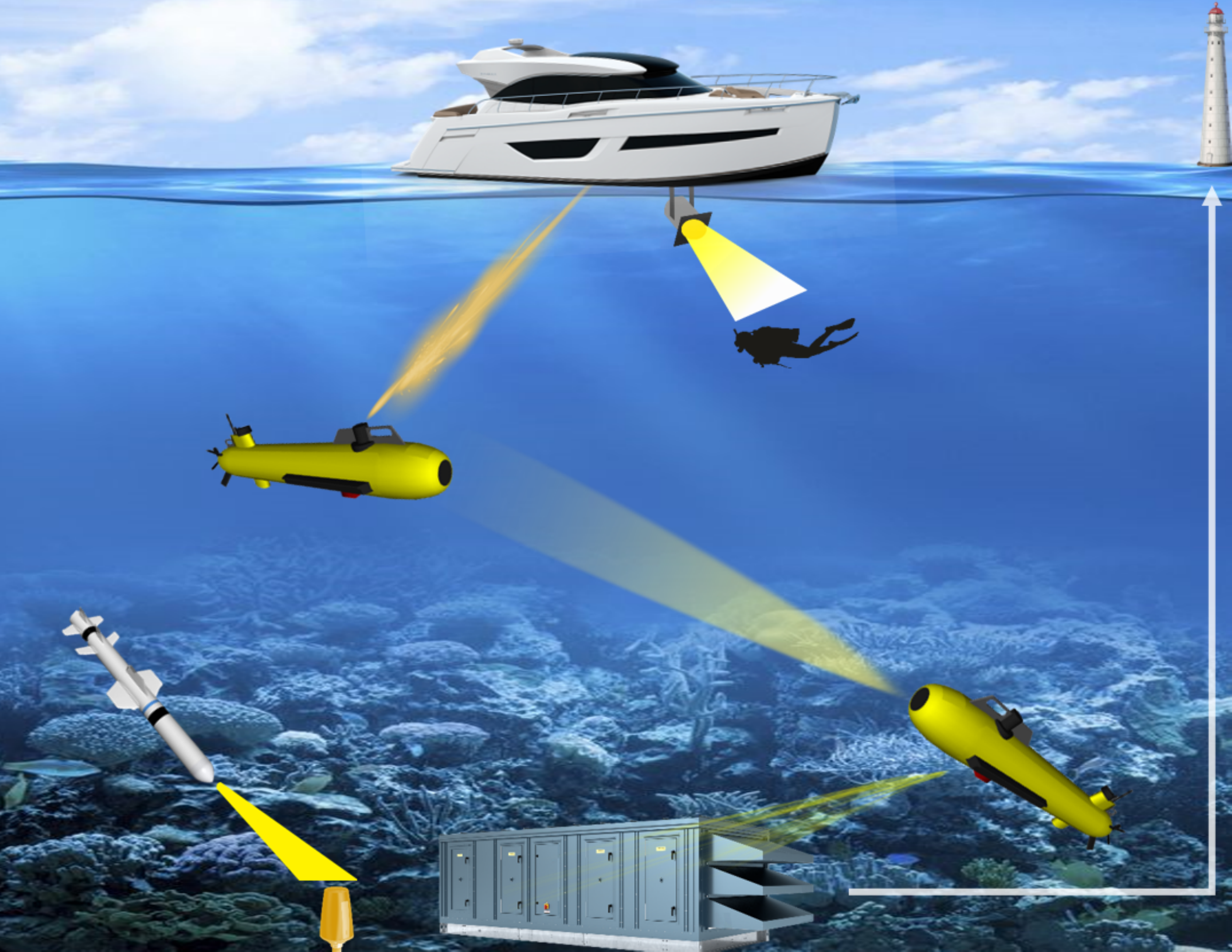}
        \put(14,45){\scriptsize{\textcolor{green}{transceiver}}}
        \put(82,6){\scriptsize{\textcolor{green}{transceiver}}}
    \put(32,13){\scriptsize{\textcolor{yellow}{Underwater data center}}}
        \end{overpic}
        \vspace{0.1cm}
        \caption{Underwater optical wireless communication network.}
        \label{UOWCNetwork} 
\end{figure}

In the future, UOWC will keep contributing on rural connection, deep ocean surveying and data storage. More infrastructures and vehicles will be connected and deployed in underwater environment to build underwater Internet of Things (IoT) to wide-area ocean networks \citep{mehedi2020systematic}, as illustrated in Fig.~\ref{UOWCNetwork}. A potential application scenario is underwater data center (UDC) as traditional data centers consume huge power and are heavy to move. UDC can cut off nearly all cooling system that reaches a power usage effectiveness (PUE) close to 1 \citep{cutler2017dunking, simon2018project,palitharathna2020multi}. 

However, underwater communication environment is relatively complicated because of the inter-symbol interference when the data rate is high, and the constant movement of the underlying optical platform caused by vibration effect, thereby adversely affecting the alignment and transmission quality. Misalignment can introduce severe pointing errors and even interrupt communication~\citep{kaushal2016underwater}. Hence, maintaining an adequate alignment is a key topic in UOWC, especially in long time underwater link~\citep{ZNBAAL:20,NZAL:21,saeed2018performance, zedini2019unified}.

To overcome these issues, performing real-time channel information through an online estimation framework to change the system behavior is an ideal method and can meet the demand. Indeed, data-driven estimation of an underwater dynamical system is an essential problem with several practical applications, specifically in control, diagnosis and monitoring. Hence, estimating the alignment angle between two underwater vehicles is a potential research direction as the alignment angle can be considered as a system state due to the inherent underwater noise and turbulence. On the other hand, the received power of LED has a strong nonlinear relationship with the alignment angle.

Extended Kalman filter (EKF) has widely been used for state estimation of nonlinear models; however, it requires linearizing the nonlinear system dynamics and result in local convergence on the mean estimate~\citep{NZZRL:20}. Solanki et al. applied EKF on estimating the angular position and angular velocity of an LED system \citep{solanki2018extended}. However, their model needs much computation and suffers from local convergence; small errors can make the systems unstable. Researchers adopt extended Luenberger observer as well if the system model can be converted into a linear one\citep{semcheddinerobust}. Mapping has been introduced to approximate the nonlinear model such as in \citep{califano2009canonical}. However, the mapping is computationally heavy to compute \citep{beineke1997comparison}, \citep{kazantzis2001discrete}.  Inspired by the idea of Luenberger's observer, researchers developed Kazantzis-Kravaris-Luenberger (KKL) observer. KKL has a good performance on both continuous and discrete-time system~\citep{kazantzis2001discrete}. However, people need to find relations between the new linear system and the previous nonlinear system to complete the coordinate transformation~\citep{poulain2008observer}. This is not always feasible in implementation.

Recently, researchers proposed to use deep learning (DL) in state estimation to improve the mapping computation easier and computationally more efficient~\citep{fadlullah2017state}. There have been few reports, but some progress, on using DL for state estimation. A deep learning (DL) based extended Luenberger observer to estimate system states has been proposed in \citep{ramos2020numerical}. This first attempt achieved good accuracy but requires labeled data for supervised learning. In addition to the difficulty of collecting data, this approach generally does not guarantee the whole inter-state mapping. However, this method has been extended to an unsupervised way~\citep{Peralezl2021Deep}. In this paper, we extended the algorithm proposed in  \citep{Peralezl2021Deep} to an LED-based optical wireless communication system to estimate the alignment angle through a deep-learning mapping.

The paper is organized as follows. In Section \ref{Model}, the LED-based optical communication model is presented, including its state-space and measurement equation. In Section \ref{sec-DP}, we formulate the deep-learning-based algorithm in which we numerically identify the learned mapping within an online framework.  In Section \ref{simulations}, simulation results are provided to illustrate the performance of the data-driven state estimation algorithm to estimate the angular position and angular velocity of the LED system through autonomous, non-autonomous, and closed-loop systems. Finally, concluding remarks are shown in Section \ref{conclusion}.

\section{LED-based Optical Communication System Model }\label{Model} 
The LED-based optical link describes a two-way communication that consists of an LED transmitter and a  photodiode receiver; each end can rotate by an angle in which it establishes and maintains LOS. 
\subsection{LED System  Dynamics}\label{Model1}
The detector's incident power can be determined based on the signal irradiance at the relative detector position. 

    \begin{figure}[!t]
        \centering
        \begin{overpic}[scale=0.26]{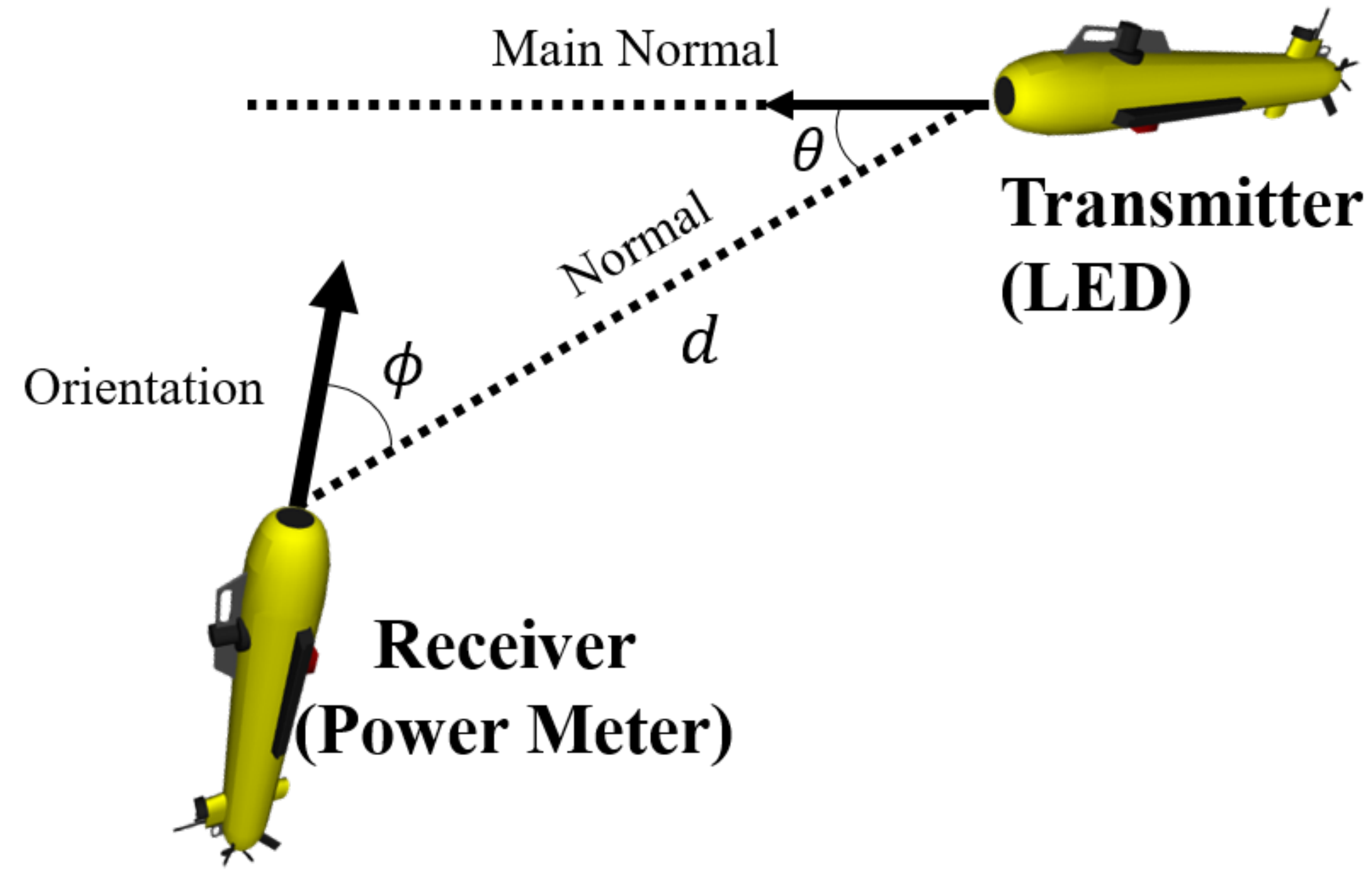}
        \end{overpic}
        \vspace{-0.1cm}
        \caption{LED communication scheme}
        \label{fig:para-demo}
    \end{figure}
The resulting luminous flux signal model is given as follows \citep{NZZRL:20}. 
\begin{equation}\label{eqa:pow-d-theta-phi}
    P_d(d, \theta, \phi) = \underbrace{\frac{a \exp{(-bd)}}{d^2} \tilde{I}_\theta}_{\mathrm{Transmitter}}\underbrace{ g(\phi)}_{\mathrm{Receiver}},
\end{equation}
where $\tilde{I}_\theta$ is the angular intensity distribution of the transmitter \citep{GPR:12,SAT:18,NZZRL:20}, $g(\phi)$ is the dependence of the received LED intensity on the incidence angle $\phi$ and takes the form of two Gaussian terms with six unknowns \citep{NZZRL:20}
\begin{equation} \label{eqa:modified-gaussian2}
    g(\phi)\!\approx \!
    a_1 \exp{\left[-\Big(\frac{\phi-b_1}{c_1}\Big)^2\right]} +
    a_2\exp{\left[-\Big(\frac{\phi+b_2}{c_2}\Big)^2\right]}.
\end{equation}
Fig.~\ref{fig:para-demo} illustrates the variables of interest, which include the transmission distance $d$, the transmission angle $\theta$, and the angle of incidence of $\phi$. All the constants values in this section are nonnegative, and their physical meaning can be found in \citep{NZZRL:20}.

From \eqref{eqa:pow-d-theta-phi}, we formulate the state space representation based on the two variables of interest $\phi\triangleq x_1$, and $\dot{\phi}\triangleq x_2$ that relate to the angles of the receiver. On the other hand, we note that practically it is not easy to move the distance $d$ ideally because it needs to move the whole robot. Besides, controlling the angular velocity of $\dot{\phi}\triangleq x_2$ is more practical. The robot alignment is performed by stabilizing the angular velocity. Since the distance $d$ cannot be adjusted easily and $\theta$ is fixed, therefore, we define the states as follows 
\begin{equation}\label{eq-1a}
{x}=\begin{bmatrix} x_1 \\ x_2  \end{bmatrix}= \begin{bmatrix}  \phi \\ \dot{\phi }\end{bmatrix}.
\end{equation}
We assume that the dynamic is slow and subject to a Gaussian process. The representation in the discrete-time domain can be written as follows
\begin{equation}\label{eq-2a}
{x_k}=\begin{bmatrix}  x_{1,k} \\ x_{2,k}\end{bmatrix}= \begin{bmatrix}  x_{1,k-1}+T_e x_{2,k-1}+w_{1,k-1} \\ x_{2,k-1}+u_{k-1}+w_{2,k-1}  \end{bmatrix},
\end{equation}
where $w_{1,k}$ and $w_{2,k}$ are the process noises which are assumed to be white independent Gaussian noises. $T_e$ is the sampling time, $u_k$ is the control input which acts on the receiver's angular velocity.

\subsection{Output Measurement Equation}
The measurement $P_{d,k}$ is expressed as
\begin{equation}\label{eq-3a}
y_k\triangleq P_{d,k}= \bar{C_p} g(x_{1,k})+w_k,
\end{equation}
where $\bar{C_p}=C_p \tilde{I}_\theta \exp{(-cd_0)}/d_0^2$ and  $g(.)$ is defined in \eqref{eqa:modified-gaussian2}.

We introduce an additional receiver on the same robot with a constant shifted angle of $\Delta \phi$ to achieve observability, as illustrated in Fig. \ref{fig1-obs1}. This shifted angle is added to account for the actual orientation of the receiver. At each movement of the transmitter platform, the states are updated according to the system dynamics. Both $\phi$ and $\bar{\phi}=\phi\pm \Delta \phi$ can be controlled to $0^{\circ}$, when $\phi$ is controlled to $0^{\circ}$~and reads the wirelessly transmitted data, its orientation is being maintained by using $\bar{\phi}$. The resulting output vector can be written as follows
\begin{equation}\label{eq-4a}
\begin{bmatrix}y_{1,k} \\ y_{2,k}
\end{bmatrix}= \bar{C_p} \begin{bmatrix} g(x_{1,k})\\ g(\underbrace{x_{1,k}\pm\Delta \phi}_{\bar{\phi}})\end{bmatrix}+w_k.
\end{equation}
    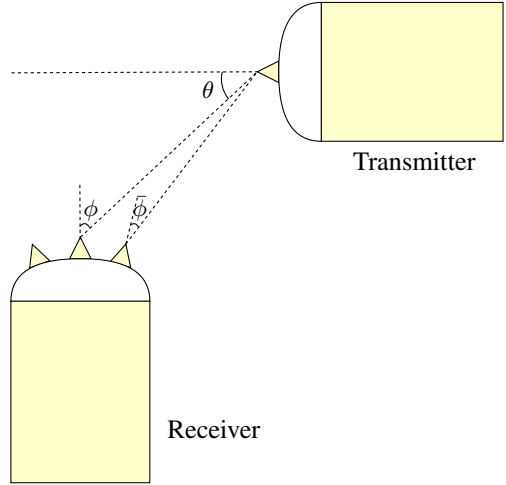
\begin{figure}[!t]
        \centering
        \begin{overpic}[scale=0.56]{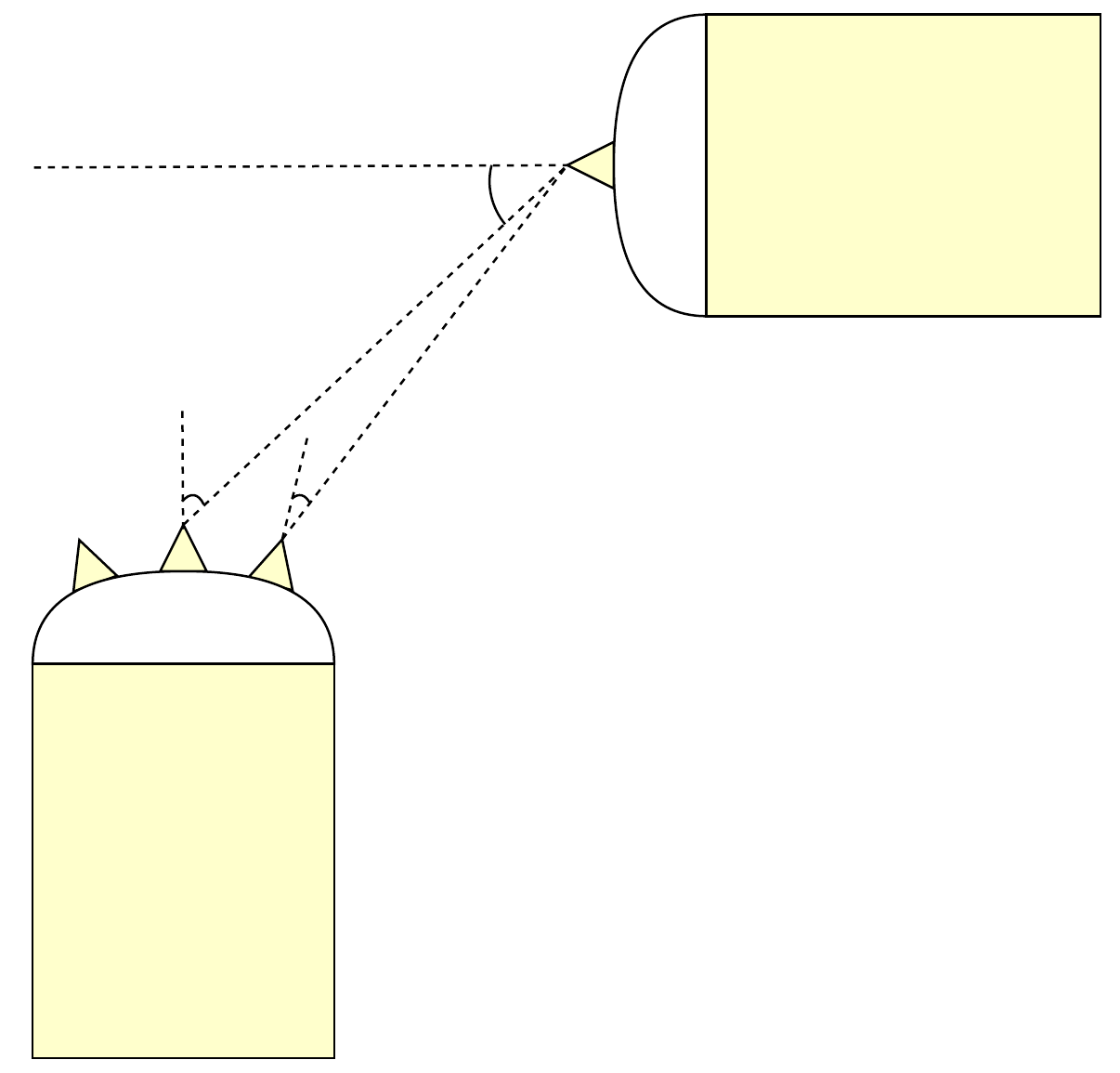} 
          \put(40,76){\footnotesize  $\theta$}
         \put(17,53){\footnotesize  $\phi$}
          \put(26.5,53){\footnotesize  $\bar{\phi}$}
      \put(68,62){ Transmitter}
        \put(32,10){ Receiver }
 \end{overpic}           
        \vspace{-0.1cm}
        \caption{Measurements of two receivers $\phi$ and $\bar{\phi}$.}
        \label{fig1-obs1}
    \end{figure}
Given the measurement, the primary goal is to estimate the angular position $x_{1,k}$ and the angular velocity $\dot{\phi}\triangleq x_{2,k}$ based on which the control $u_k$ is designed, to drive $x_{2,k}$ towards zero, which corresponds to the maximum light intensity's orientation. In this paper, we rely on the deep learning based-observer design results \citep{Peralezl2021Deep,BeA:19}, to online estimate $x_{1,k}$ and $x_{2,k}$ from the knowledge of a sequence of the past and current values of the input $u_k$ and output $y_k$.

The next section provides the design and algorithm of the deep learning observer-based reference tracking control.

\section{Deep learning-based estimation algorithm}\label{sec-DP}
We consider the following non-autonomous system
\begin{equation}\label{equat-1a}
    \left\{\begin{array}{l}x_{k+1}=f\left(x_{k}, u_{k}\right) \\ y_{k}=\ell\left(x_{k}\right)
    \end{array} \right.
\end{equation}
where $x\in \Rn$ is the state, $u\in \Rm$ is the input and $y\in \Rp$ is the output. $f$ and $\ell$ are suitable nonlinear functions.

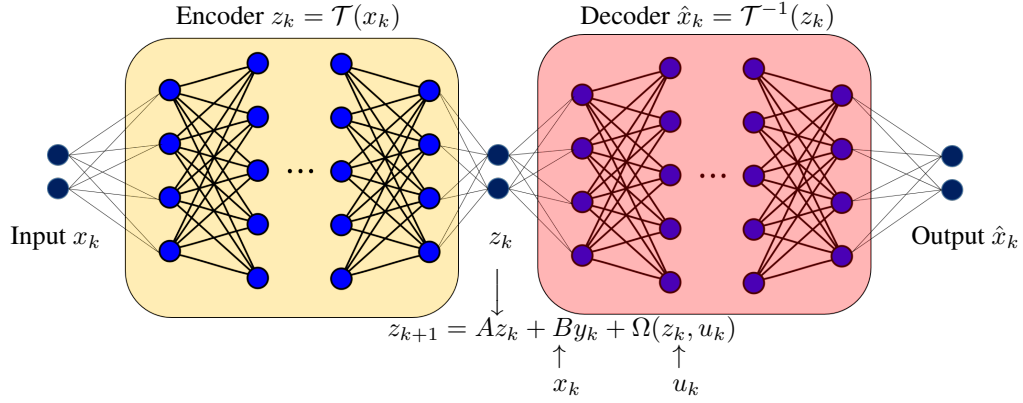
\begin{figure*}[!t]
        \centering
        \begin{overpic}[scale=0.18]{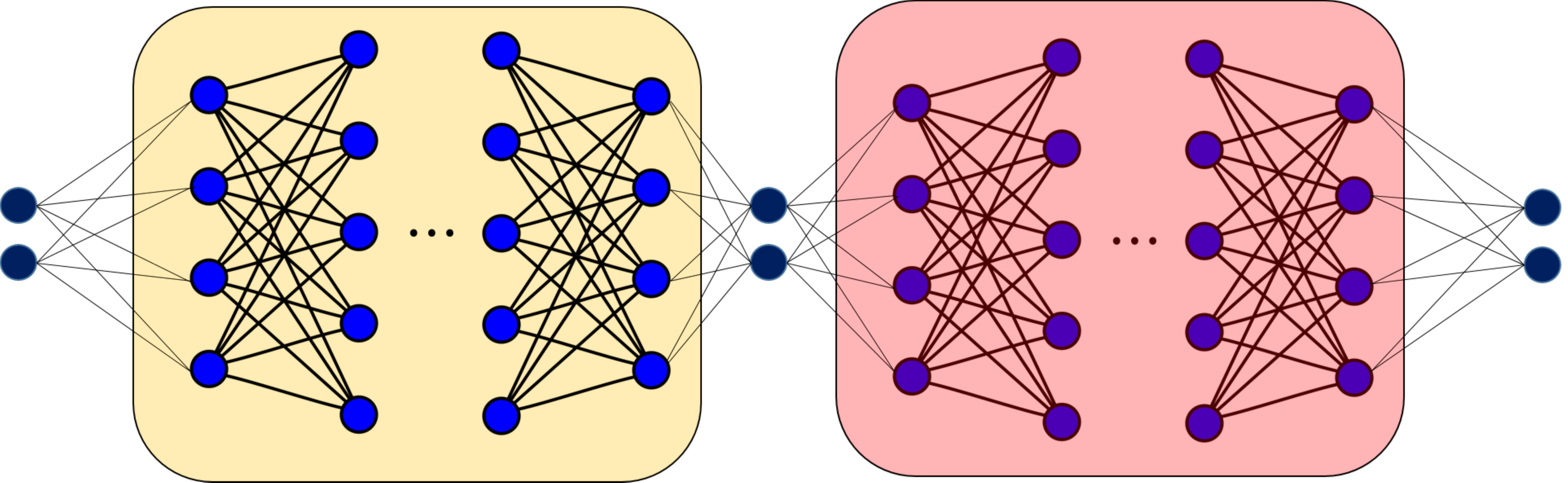}
         \put(-4,8){Input $x_k$}
          \put(14,32){Encoder $z_k=\mathcal{T}(x_k)$}
             \put(58,32){Decoder $\hat{x}_k=\mathcal{T}^{-1}(z_k)$}
               \put(94,8){Output $\hat{x}_k$}
        \put(48,8){$z_k$}
        \put(48,2){$\Big\downarrow$}
          \put(37,-2){$z_{k+1}=Az_{k}+B y_{k}+\Omega(z_k,u_k)$}
                  \put(55,-5){$\uparrow$}
                \put(55,-8){$x_k$}
                \put(68,-5){$\uparrow$}
                \put(68,-8){$u_k$}
        \end{overpic}\vspace{1cm}
        \caption{Structure of the deep auto-encoder network model to identify $\mathcal{T}$ and $\mathcal{T}^{-1}$.}
        \label{D-auto}
        \vspace{0.25cm}
\end{figure*}

\subsection{Preliminaries}\label{subsec-1}
Throughout the paper, we assume that $f$ and $\ell$ satisfy the following assumptions.
\begin{ass}\label{ass-1}
$f$ is invertible and $f^{-1}$ and $\ell$ are of class $\C^1$ and globally Lipschitz.
\end{ass}

\begin{ass}\label{ass-2}
For all $(x_1,x_2)\in\mathcal{X}\subset\Rn$ of system \eqref{equat-1a} with input $u$, if $x_1\neq x_2$, then there exists a positive integer $j$ such that $\ell\left(f^{-j}(x_1)\right)\neq \ell(f^{-j}(x_2))$.
\end{ass}
For brevity, we require system \eqref{equat-1a} to be reversible in time and thus make Assumption \ref{ass-1}. Assumption \ref{ass-2} implies a backward distinguishability hypothesis of the state function \eqref{equat-1a} to ensure sufficient conditions of the existence, injectivity and uniqueness of a map $\mathcal{T}$.
\begin{lemma}\label{lem1}\citep{Peralezl2021Deep}
Suppose that assumptions \ref{ass-1} and \ref{ass-2} hold and for any constant input $\bar{u}$, there exists a map $\mathcal{T}\!:\mathcal{X}\rightarrow \Rq$ for almost any controllable pair $(A,B)$ of dimension $q=p(n+1)$ with $A$ Hurwitz that ensures
\begin{equation}\label{equat1a}
 \mathcal{T}(f(x,\bar{u}))=A\mathcal{T}(x)+B\ell(x), \quad \forall x\in \mathcal{X}, 
\end{equation}
and a pseudo-inverse $\mathcal{T}^{-1}$ such that the following system 
\begin{equation}\label{equat2a}
    \left\{\begin{array}{l}z_{k+1}=A z_{k} +B y_{k} \\ \hat{x}_{k}=\mathcal{T}^{-1}(z_k)
    \end{array} \right.
\end{equation}
is an observer for \eqref{equat-1a} for a constant input $\bar{u}$. Then, the unique solution of \eqref{equat1a} is given as follows
\begin{equation}\label{equat3a}
\mathcal{T}(x)=\sum_{j=0}^{+\infty}A^jB\ell\left(f^{-(j+1)}(x,\bar{u})\right).
\end{equation}
\end{lemma}

\begin{lemma}\citep{Peralezl2021Deep} Let assumptions \ref{ass-1} and \ref{ass-2} hold for a constant input $\bar{u}$. Assume that $A$ can be obtained such that $\mathcal{T}$ and $\mathcal{T}^{-1}$ in lemma \ref{lem1} satisfy
\begin{equation*}
 \abs{\Omega(z_1,u)-\Omega(z_2,u)} \leqslant \lambda_u \abs{z_1-z_2},
\end{equation*}
where $\abs{.}$ is the Euclidean norm and \begin{equation*}
\Omega(z_k,u_k)=\mathcal{T}\left(f\left(\mathcal{T}^{-1}(z_k),u_k\right)\right)-\mathcal{T}\left(f\left(\mathcal{T}^{-1}(z_k),\bar{u}\right)\right).
\end{equation*}
Then, for all $u\in\mathcal{U}$ such that $\rho(A+\lambda_u I)<1$, the following system
\begin{equation}\label{equatcol1a}
    \left\{\begin{array}{l}z_{k+1}=A z_{k} +B y_{k}+\Omega(z_k,u_k) \\ \hat{x}_{k}=\mathcal{T}^{-1}(z_k)
    \end{array} \right.
\end{equation}
is an observer for \eqref{equat-1a}.
\end{lemma}
\begin{remark}
An analytic expression of $\mathcal{T}$ (i.e., the existence and injectivity of $\mathcal{T}$) of this class of linear LED non-autonomous dynamics with exponential output \eqref{eq-2a}-\eqref{eq-4a} can be proven by relying on the resolution of a time-varying PDE that provides solutions of which transform the dynamics \eqref{eq-2a}-\eqref{eq-4a} into linear asymptotically stable ones (see, for instance \citep{BeA:19}). However, it is essential to mention that an explicit close form expression of $\mathcal{T}^{-1}$ is usually challenging to obtain, and it is not always straightforward to determine the function $\mathcal{T}$ derived from \eqref{equat3a}. 
\end{remark}

Our next objective is to numerically identify the injective mapping $\mathcal{T}$ and its left inverse $\mathcal{T}^{-1}$.

\subsection{Online Estimation of $\mathcal{T}$ and $\mathcal{T}^{-1}$}\label{subsec-3}
A deep auto-encoder network is used to identify the mapping of $\mathcal{T}$ that satisfies \eqref{equat1a} along with $\mathcal{T}^{-1}$ for non-autonomous system, as shown in Fig.~\ref{D-auto}. Hence, two loss functions are computed during the training phase \citep{Peralezl2021Deep}. The first loss function minimizes the trajectory dynamic to identify a latent space $\mathcal{T}$ as follows
\begin{equation}\label{cost1}
\mathcal{L}_{\mbox{\scriptsize dyn}}=\norm{\mathcal{T}(x_{k+1})-A\mathcal{T}(x_k)+B\ell(x_k)+\bar{\Omega}(x_k,u_k)},
\end{equation}
where 
\begin{equation*}
\bar{\Omega}(x_k,u_k)=\mathcal{T}\left(f(x_k,u_k)\right)-\mathcal{T}\left(f(x_k,\bar{u})\right).
\end{equation*}

The second loss function uses a reconstruction loss of the auto-encoder to learn $\mathcal{T}^{-1}$ such that $\hat{x}$ is recovered. The reconstruction cost is given by
\begin{equation}\label{cost2}
\mathcal{L}_{\mbox{\scriptsize recon}}=\norm{x_{k}-\mathcal{T}^{-1}\left(\mathcal{T}(x_k)\right)},
\end{equation}
where $\norm{.}$ is the mean squared-error.

The LED-based optical model \eqref{eq-2a}-\eqref{eq-4a} is trained by minimizing the two loss functions \eqref{cost1} and \eqref{cost2} on the dataset $\mathcal{D}=\{x_k, x_{k+1}, \bar{u}\}$ where $x_k$ and $x_{k+1}$ are generated from a uniform random distribution on $\mathcal{X}$ and from the LED model dynamic \eqref{eq-2a}-\eqref{eq-4a}, respectively. For some constant value $\bar{u}$ of the input, we first learn the mapping $\mathcal{T}$ and $\mathcal{T}^{-1}$, then we apply a small enough input during evaluation which can be seen as a disturbance. Algorithm~\ref{RoughStepsAlgorithm} summarizes the steps of the deep learning based-observer design.
\vspace{0.25cm}
\begin{algorithm}
\SetAlgoLined
\caption{Online estimation of $\mathcal{T}$ and $\mathcal{T}^{-1}$ }
\label{RoughStepsAlgorithm}
\KwData {Choose $A$ and $B$\\ Initialize a data set ${(x_k,x_{k+1}, \bar{u})}$ and  $\mathcal{T}$ randomly}
\While{$k \leqslant max\_iter$}{
  Compute $z_k=\mathcal{T}(x_k)$\\
  \While{$i \leqslant$ dimension of $z$}{
    $b_{i}=\frac{1}{D\left(z_{i, k}\right)}$
  }{$\mathcal{T}(x)\leftarrow\operatorname{diag}\left(\left[b_{1}, b_{2}, \cdots\right]\right) \mathcal{T}(x)$ \\
    minimize $\mathcal{L}_{\mbox{\scriptsize dyn}}=\norm{\mathcal{T}(x_{k+1})-A\mathcal{T}(x_k)+B\ell(x_k)+\bar{\Omega}(x_k,u_k)}$
  }
}
\While{$k \leqslant max\_iter$}{
minimize $\mathcal{L}_{\mbox{\scriptsize recon}}=\norm{x_{k}-\mathcal{T}^{-1}\left(\mathcal{T}(x_k)\right)}$
}
\end{algorithm}
\vspace{0.25cm}

\begin{remark}
Note that the white Gaussian process noises can be embedded in the additive input signal. Hence, the learned mappings $\mathcal{T}$ and $\mathcal{T}^{-1}$ hold for a small excitation signal as the contraction property \eqref{equatcol1a} remains still satisfied.
\end{remark}

\begin{figure*}[!t]
        \centering
        \begin{overpic}[scale=1.82]{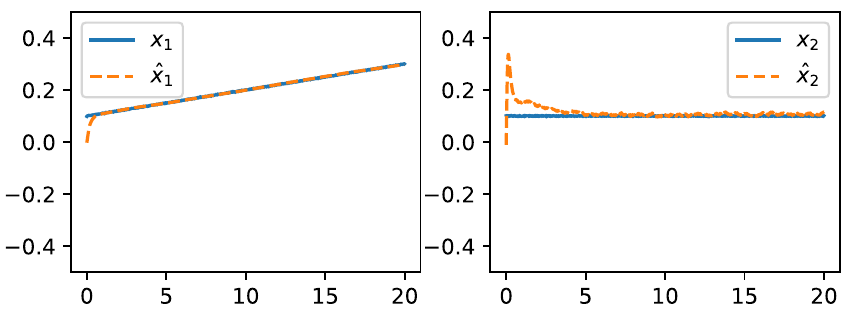}
        \put(0,9){\begin{rotate}{90} Angular position [rad] \end{rotate}}
      \put(25,-2){ Time [s]}
    \put(101,30){\begin{rotate}{-90} Angular velocity [rad/s] \end{rotate}}
      \put(75,-2){ Time [s]}
        \end{overpic}
        \vspace{0.2cm}
        \caption{Open-loop case with $u=0$: true states $x_1$ and $x_2$ and their estimated states $\hat{x}_1$ and $\hat{x}_2$, respectively.}
        \label{Performanceu0Estimation}
\end{figure*}

\begin{figure*}[!t]
        \centering
        \begin{overpic}[scale=1.82]{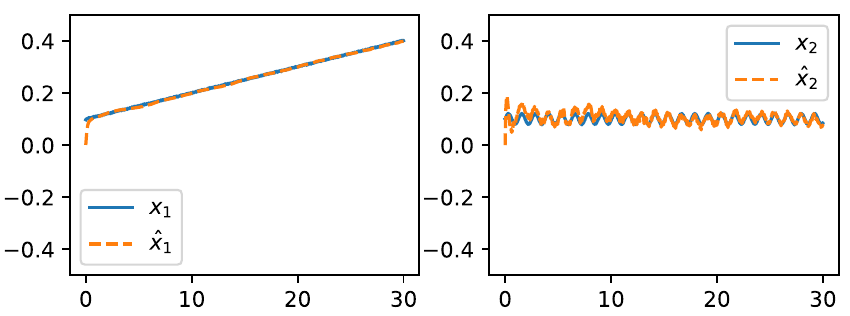}
                       \put(0,9){\begin{rotate}{90} Angular position [rad] \end{rotate}}
      \put(25,-2){ Time [s]}
    \put(101,30){\begin{rotate}{-90} Angular velocity [rad/s] \end{rotate}}
      \put(75,-2){ Time [s]}
        \end{overpic}
        \vspace{0.2cm}
        \caption{Non-autonomous case with $u\!=\! 0.1\cos\left({5t\times0.01}\right)$: true states $x_1$ and $x_2$ and their estimated states $\hat{x}_1$ and $\hat{x}_2$, respectively.}
        \label{Performanceu0.1cosEstimation} 
\end{figure*}

\begin{figure*}[!t]
        \centering
        \begin{overpic}[scale=1.82]{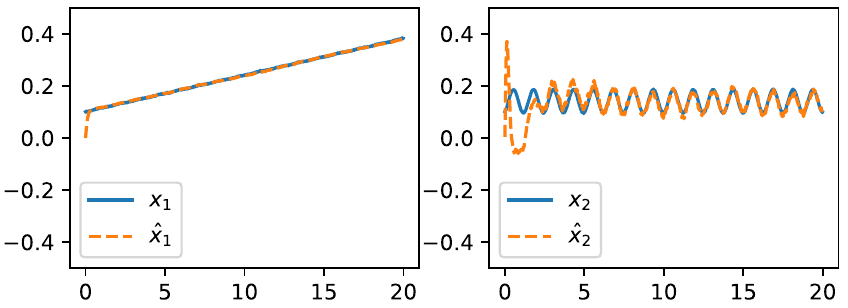}
                       \put(0,9){\begin{rotate}{90} Angular position [rad] \end{rotate}}
      \put(25,-2){ Time [s]}
    \put(101,30){\begin{rotate}{-90} Angular velocity [rad/s] \end{rotate}}
      \put(75,-2){ Time [s]}
        \end{overpic}
        \vspace{0.2cm}
        \caption{Non-autonomous case with $u\!=\!0.1\cos \left({5t\times0.01}\right)+0.2\sin\left({5t \times 0.01}\right)$: true states $x_1$ and $x_2$ and their estimated states $\hat{x}_1$ and $\hat{x}_2$, respectively.}
        \label{Performanceu0.1cos0.2sinEstimation} 
\end{figure*}
 
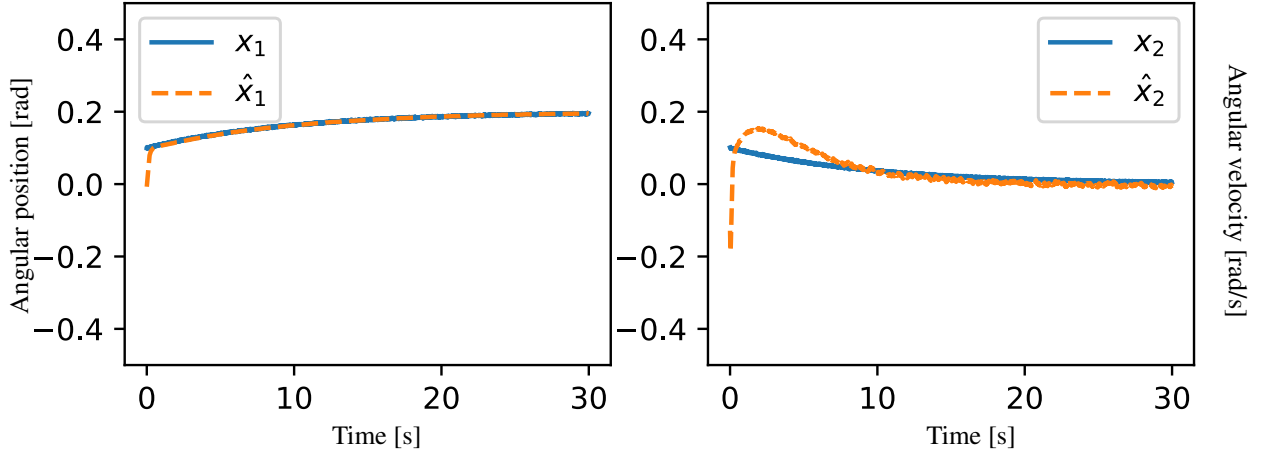
\begin{figure*}[!t]
        \centering
        \begin{overpic}[scale=1.82]{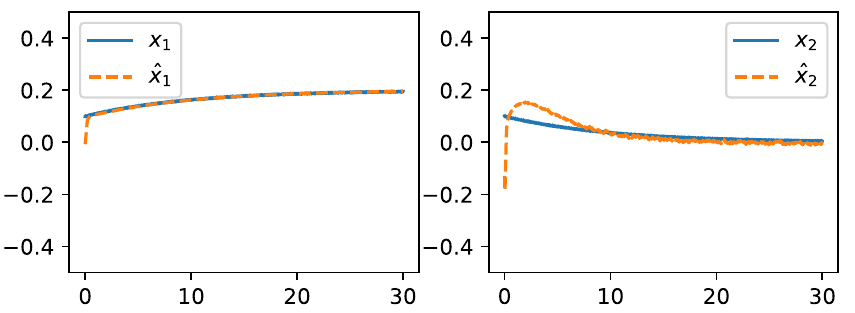}
                       \put(0,9){\begin{rotate}{90} Angular position [rad] \end{rotate}}
      \put(25,-2){ Time [s]}
        \put(101,30){\begin{rotate}{-90} Angular velocity [rad/s] \end{rotate}}
      \put(75,-2){ Time [s]}
        \end{overpic}
        \vspace{0.2cm}
        \caption{Closed-loop setting with $u=-0.001{\hat{x}_2}$: true states $x_1$ and $x_2$ and their estimated states $\hat{x}_1$ and $\hat{x}_2$, respectively.}
        \label{Performanceu-0.001Estimation} 
\end{figure*}

\begin{table*}[!t]
\vspace{0.5cm}
    \centering
        \caption{RMSE: state and output errors performance estimation.}
    \begin{tabular}{cccccc}
    \toprule			
Controllers &~RMSE~	&	$~x_1-\hat{x}_1$[{\si rad}]~	&	~$~x_2-\hat{x}_2$[{\si rad/s}]~	&	~$y_1-\hat{y}_1$[{\si W}]~	&	$y_2-\hat{y}_2$[{\si W}]~	\\	\midrule
Open-loop (OL)& $u=0$	&	$5.2\times10^{-3}$	&	$19.2\times10^{-3}$	&	$8.8\times10^{-3}$	&	$13.6\times10^{-3}$	\\	
Input function $\mbox{IF}^1$ & $u = 0.1\cos \left({5t \times 0.01} \right)$	& $7.2\times10^{-3}$	& $51.8\times10^{-3}$	&	$11.6\times10^{-3}$	&	$19.3\times10^{-3}$	\\	
Input function $\mbox{IF}^2$  & $u = 0.1\cos \left( {5t \times 0.01} \right) + 0.2\sin \left( {5t \times 0.01} \right)$ 	&	$5.7\times10^{-3}$	&	$50.3\times10^{-3}$
& $1.73\times10^{-3}$	&	$2.5\times10^{-3}$	
\\	
Closed-loop (CL)  & $u =-0.001 \hat{x}_2$ 	&	$6.3\times10^{-3}$	&	$22\times10^{-3}$	& $8.4\times10^{-3}$	&	$16.7\times10^{-3}$ \\	\bottomrule
\end{tabular}\label{tab:rmse1}
\vspace{0.2cm}
\end{table*}

\section{Simulation Results}\label{simulations}
In this section, we present numerical simulation results that illustrate the performance of the proposed online estimation methodology. The simulations are carried out by using Python. We employ synthetic data generated from the LED-based optical model \eqref{eq-2a}-\eqref{eq-4a}. We apply the proposed deep learning observer algorithm to minimize the two-loss functions \eqref{cost1} and \eqref{cost2} on the training data set.  We consider three scenarios for the new inputs: the first deals with the open-loop case, the second considers the non-autonomous case with two different external inputs, and the last is built from the closed-loop setting. We first identify the learning mapping $\mathcal{T}$ and $\mathcal{T}^{-1}$ in open-loop setting (i.e. $\bar{u}=0$). Then, we predict the LED states using the learned model with new inputs and initial conditions. Finally, we compare the prediction results with the reference solution, which is derived by solving the exact LED system  \eqref{eq-2a}-\eqref{eq-4a} with the same new inputs. We conduct the deep neural network training using Adam optimizer through dense neural networks with the open-source Pytorch library. The LED model is trained with $2\times 10^5$ data trajectories randomly sampled with a uniform distribution on the domain $\mathcal{X}=[-0.5, 0.5]\times[-0.5, 0.5]$. We consider the eigenvalues for the observer, corresponding to $A\!=\!\diag[1-T_e, 1-2T_e,1-4T_e,1-6T_e, 1-8T_e,1-10T_e]$ and matrix $B$ is given by $B=\mbox{ones}(6,2)$ in our discrete framework. The measurement power is corrupted with zero mean Gaussian noise of $0.001$ variance and the shifted angle $\Delta\phi=6^\circ$. The deep learning algorithm uses activation function $\tanh(.)$ and consists of one hidden layer for both networks with $500$ nodes. Overall, the hyperparameters of the DL algorithm are chosen to ensure a good compromise between the estimation performance and minimum loss function. 

\begin{figure*}[!t]
        \centering
        \begin{overpic}[scale=1.82]{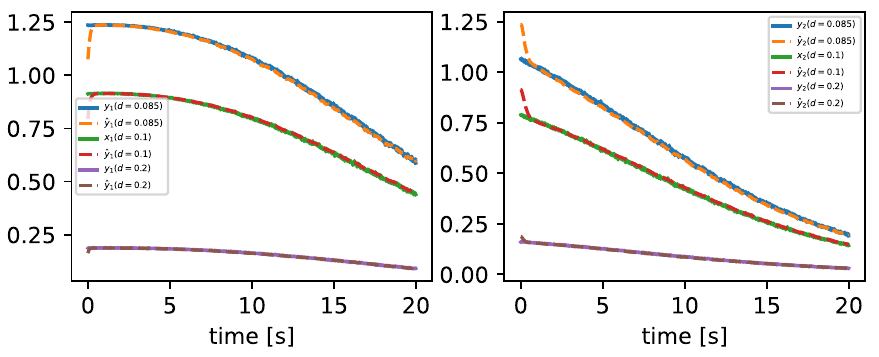}
\put(-1,12){\begin{rotate}{90} Power signal strength [W] \end{rotate}}
\put(101,36){\begin{rotate}{-90} Power signal strength [W] \end{rotate}}
        \end{overpic}
        \vspace{-0.2cm}
        \caption{Output prediction results along with their reference solutions when $u = 0.1\cos \left({5t\times0.01}\right)$: effect of the power signal strength to the distance between the receiver and the transmitter.}
     \label{Performanceu0.1cosSensitivity} 
\end{figure*}

\begin{figure*}[!t]
        \centering
        \begin{overpic}[scale=1.82]{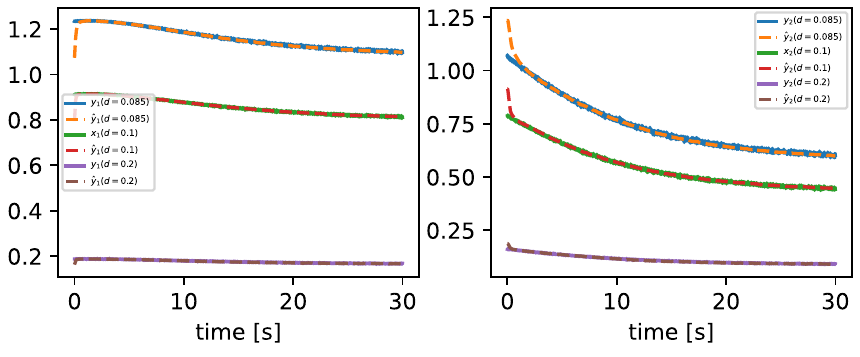}
\put(-1,12){\begin{rotate}{90} Power signal strength [W] \end{rotate}}
\put(101,36){\begin{rotate}{-90} Power signal strength [W] \end{rotate}}
        \end{overpic}
        \vspace{-0.2cm}
\caption{Output prediction results along with their reference solutions when $u\!=\!0.1\cos\left({5t\times0.01}\right)+0.2\sin \left({5t\times0.01}\right)$: effect of the power signal strength to the distance between the receiver and the transmitter.} 
    \label{Performanceu0.1cos0.2sinSensitivity} 
\end{figure*}
\begin{figure*}[!t]
        \centering
        \begin{overpic}[scale=1.82]{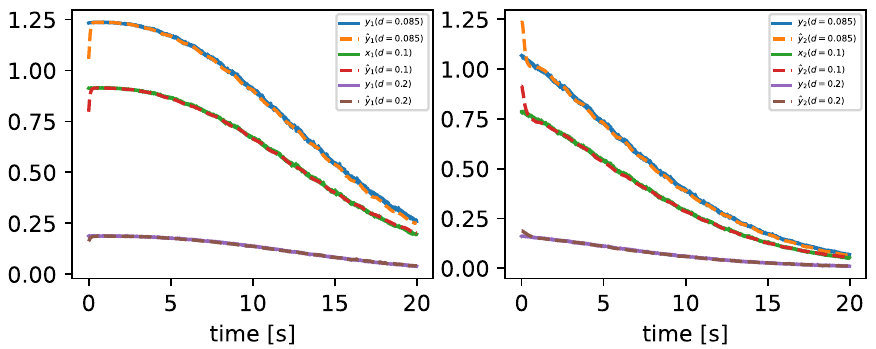}
\put(-1,12){\begin{rotate}{90} Power signal strength [W] \end{rotate}}
\put(101,36){\begin{rotate}{-90} Power signal strength [W] \end{rotate}}
        \end{overpic}
        \vspace{-0.2cm}
\caption{Output prediction results along with their reference solutions when $u\!=\!- 0.001{\hat{x}_2}$: effect of the power signal strength to the distance between the receiver and the transmitter.}
\label{Performanceu-0.001Sensitivity}
\vspace{0.5cm}
\end{figure*}

\subsection{LED States Estimation}\label{subsec}

\subsubsection{Open-loop simulations}
We firstly perform open-loop simulation results to illustrate the importance of estimating the mapping $\mathcal{T}$ and $\mathcal{T}^{-1}$ in the unforced LED model. Fig.~\ref{Performanceu0Estimation} shows the open-loop simulation results. $u=0$ and the root-mean-square error (RMSE) is shown in Table~\ref{tab:rmse1}. It can be seen that the observer exhibits good performance with the reference solution. Hence, the results illustrate the ability of the proposed method to identify the mapping $\mathcal{T}$ and $\mathcal{T}^{-1}$.

\subsubsection{Non-autonomous simulations}
For comparison purposes, we perform simulation results  for two non-autonomous cases $\mbox{IF}^1$ and $\mbox{IF}^2$ with $u\!=\!0.1\cos\left({5t\times0.01}\right)$ and  $u\!=\!0.1\cos\left({5t\times0.01}\right)+ 0.2\sin\left({5t\times0.01}\right)$ along with the reference solution, as illustrated in Figs.~\ref{Performanceu0.1cosEstimation} and \ref{Performanceu0.1cos0.2sinEstimation}, respectively.  Table~\ref{tab:rmse1} shows the corresponding RMSE for both input functions $\mbox{IF}^1$ and $\mbox{IF}^2$. Hence, we observe good agreement of the angular position and velocity with the reference solution for relatively small enough input signal. From Table~\ref{tab:rmse1}, it is worth noting that the RMSE increases slightly for both external inputs compared to the open-loop case, however the overall results exhibit good performances. Hence, one can conclude that the learned mappings can be used to estimate and predict system behavior for adequate external inputs.

\subsubsection{Closed-loop simulations}\label{CL}
For further performance analysis, we perform a closed-loop test based on the estimated angular velocity. We set the angular reference position to $0.2${\si rad} and the angular velocity to zero. Fig.~\ref{Performanceu-0.001Estimation} illustrates the estimated angular position and velocity states of the LED and the output power prediction results under a closed-loop controller along their reference solutions. Table~\ref{tab:rmse1} also provides the corresponding RMSE.
We observe good performance with the true states.

\subsection{Sensitivity Analysis}
We have conducted simulations in the loop with the inputs signal and the closed-loop controller to examine the performance of the optical communication link when the distance between the receiver and the transmitter increases. As illustrated in Figs. \ref{Performanceu0.1cosSensitivity}, \ref{Performanceu0.1cos0.2sinSensitivity}, and \ref{Performanceu-0.001Sensitivity}, we observe that the signal strength significantly decreases in both  excitation input and closed-loop signals when the distance between the transmitter and the receiver increases. Furthermore, Table~\ref{tab:rmse2} shows the RMSE of the state estimation and output power prediction errors performance with different link distances and control inputs and confirms that the receiver signal strength continuously decreases with the link distances over the prediction time.

\begin{table*}[!t]
    \centering
        \caption{RMSE: state and output errors performance estimation with different link distances and controllers.}
    \begin{tabular}{cccccc}
    \toprule			
 ~Controller~	& Distance &	$~x_1-\hat{x}_1$[{\si rad}]~	&	~$~x_2-\hat{x}_2$[{\si rad/s}]~	&	~$y_1-\hat{y}_1$[{\si W}]~	&	$y_2-\hat{y}_2$[{\si W}]~	\\	\midrule
$u = 0.1\cos \left({5t \times 0.01} \right)$	& $d=0.085$ &  $7.4\times10^{-3}$ 	 &  $51.7\times10^{-3}$	&	$11.6\times10^{-3}$	&	$19.3\times10^{-3}$	\\	
&  $d=0.1$ 	&	$6.9\times10^{-3}$	&	$45.6\times10^{-3}$
& $7.8\times10^{-3}$	&	$13.4\times10^{-3}$	
\\	
  &  $d=0.2$ 	&	$5.7\times10^{-3}$	&	$50.3\times10^{-3}$	& $1.73\times10^{-3}$	&	$2.54\times10^{-3}$ \\	\midrule
     ~~~$u = 0.1\cos \left( {5t \times 0.01} \right) + 0.2\sin \left( {5t \times 0.01} \right)$~~~~~~	&   $d=0.085$ &	$6.64\times10^{-3}$	&	$40.1\times10^{-3}$ &    $13.0\times10^{-3}$ &    $15.6\times10^{-3}$ \\
&  $d=0.1$	&	$5.7\times10^{-3}$	&	$31.5\times10^{-3}$
& $6.8\times10^{-3}$ & $10.9\times10^{-3}$	
\\	
  & $d=0.2$ 	&	$7.3\times10^{-3}$	&	$24.7\times10^{-3}$	& $2.2\times10^{-3}$	&	$3.1\times10^{-3}$ \\	\midrule
 $u =-0.001 \hat{x}_2$	& $d=0.085$ &  $6.3\times10^{-3}$	& $22.0\times10^{-3}$	&	$8.4\times10^{-3}$	&	$16.7\times10^{-3}$	\\	
&  $d=0.1$ 	&	$4.9\times10^{-3}$	&	$43.3\times10^{-3}$
& $5.1\times10^{-3}$	&	$9.5\times10^{-3}$	
\\	
  &  $d=0.2$ 	&	$5.4\times10^{-3}$	&	$30.9\times10^{-3}$	& $1.6\times10^{-3}$	&	$2.5\times10^{-3}$ \\	\bottomrule
\end{tabular}
    \label{tab:rmse2}
    \vspace{0.5cm}
\end{table*}

\section{Conclusion}\label{conclusion}
In this paper, we have leveraged the power of deep learning-based observer design to estimate the LED state variables and analyze the output power prediction under measurement noise on the underwater optical communication channel. Indeed, we presented a numerical method for constructing the mapping, which drives the LED discrete-time nonlinear system into a stable linear system. We then designed an observer to estimate the angular position and angular velocity of the LED system. Using the learned mapping for constant input setting (for instance, open-loop setting) and measurement noises, we have shown that the proposed deep learning framework can identify the LED states for small enough inputs signal, including a closed-loop control. 

\section*{Acknowledgment}
The authors would like to thank Dr. Madiha Nadri and her team for their interesting discussions. 

\bibliographystyle{unsrtnat}
\def\BibTeX{{\rm B\kern-.05em{\sc i\kern-.025em b}\kern-.08em
    T\kern-.1667em\lower.7ex\hbox{E}\kern-.125emX}}
\balance
\bibliography{ref.bib}

\begin{thebibliography}{33}
\providecommand{\natexlab}[1]{#1}
\providecommand{\url}[1]{\texttt{#1}}
\expandafter\ifx\csname urlstyle\endcsname\relax
  \providecommand{\doi}[1]{doi: #1}\else
  \providecommand{\doi}{doi: \begingroup \urlstyle{rm}\Url}\fi

\bibitem[Schirripa~Spagnolo et~al.(2020)Schirripa~Spagnolo, Cozzella, and
  Leccese]{schirripa2020underwater}
Giuseppe Schirripa~Spagnolo, Lorenzo Cozzella, and Fabio Leccese.
\newblock Underwater optical wireless communications: Overview.
\newblock \emph{Sensors}, 20\penalty0 (8):\penalty0 2261, 2020.

\bibitem[Gkoura et~al.(2014)Gkoura, Nistazakis, Vavoulas, Tsigopoulos, and
  Tombras]{gkoura2014underwater}
LK~Gkoura, HE~Nistazakis, A~Vavoulas, AD~Tsigopoulos, and GS~Tombras.
\newblock Underwater optical wireless communications possibilities
  disadvantages and possible solutions.
\newblock In \emph{6th International Conference from Scientific Computing to
  Computational Engineering IC-SCCE}, pages 9--12, 2014.

\bibitem[Cossu(2019)]{cossu2019recent}
Giulio Cossu.
\newblock Recent achievements on underwater optical wireless communication.
\newblock \emph{Chinese Optics Letters}, 17\penalty0 (10):\penalty0 100009,
  2019.

\bibitem[Tian et~al.(2013)Tian, Zhang, and Tan]{tian2013design}
Bin Tian, Feitian Zhang, and Xiaobo Tan.
\newblock Design and development of an {LED}-based optical communication system
  for autonomous underwater robots.
\newblock In \emph{2013 IEEE/ASME International Conference on Advanced
  Intelligent Mechatronics}, pages 1558--1563. IEEE, 2013.

\bibitem[Wang et~al.(2018)Wang, Li, and Xu]{wang2018cost}
Peilin Wang, Chao Li, and Zhengyuan Xu.
\newblock A cost-efficient real-time 25 mb/s system for {LED-UOWC}: design,
  channel coding, {FPGA} implementation, and characterization.
\newblock \emph{Journal of Lightwave Technology}, 36\penalty0 (13):\penalty0
  2627--2637, 2018.

\bibitem[Arvanitakis et~al.(2020)Arvanitakis, Bian, McKendry, Cheng, Xie, He,
  Yang, Islim, Purwita, Gu, et~al.]{arvanitakis2020gb}
Georgios~N Arvanitakis, Rui Bian, Jonathan~JD McKendry, Chen Cheng, Enyuan Xie,
  Xiangyu He, Gang Yang, Mohamed~Sufyan Islim, Ardimas~A Purwita, Erdan Gu,
  et~al.
\newblock Gb/s underwater wireless optical communications using
  series-connected gan micro-{LED} arrays.
\newblock \emph{IEEE Photonics Journal}, 12\penalty0 (2), 2020.

\bibitem[Son et~al.(2018)Son, Kang, Nhat, Kim, and Choi]{son2018study}
Hyun-Joong Son, Jin-Il Kang, Thieu Quang~Minh Nhat, Seo~Kang Kim, and
  Hyeung-Sik Choi.
\newblock Study on underwater optical communication system for video
  transmission.
\newblock \emph{Journal of Ocean Engineering and Technology}, 32\penalty0
  (2):\penalty0 143--150, 2018.

\bibitem[Sticklus et~al.(2018)Sticklus, Hoeher, and
  R{\"o}ttgers]{sticklus2018optical}
Jan Sticklus, Peter~Adam Hoeher, and R{\"u}diger R{\"o}ttgers.
\newblock Optical underwater communication: The potential of using converted
  green {LEDs} in coastal waters.
\newblock \emph{IEEE Journal of Oceanic Engineering}, 44\penalty0 (2):\penalty0
  535--547, 2018.

\bibitem[Jiang(2018)]{jiang2018overview}
Shengming Jiang.
\newblock Overview of underwater acoustic communication.
\newblock In \emph{Wireless networking principles: from terrestrial to
  underwater acoustic}, pages 233--244. Springer, 2018.

\bibitem[Kumara and Vatsb(2021)]{kumara2021underwater}
Suresh Kumara and Chanderkant Vatsb.
\newblock Underwater communication: A detailed review.
\newblock In \emph{CEUR Workshop Proceedings}, 2021.

\bibitem[Saini et~al.(2017)Saini, Singh, and Sinha]{saini2017path}
Preeti Saini, Rishi~Pal Singh, and Adwitiya Sinha.
\newblock Path loss analysis of rf waves for underwater wireless sensor
  networks.
\newblock In \emph{2017 International Conference on Computing and Communication
  Technologies for Smart Nation (IC3TSN)}, pages 104--108. IEEE, 2017.

\bibitem[Mehedi et~al.(2020)Mehedi, Hasan, Sadiq, Akhtar, and
  Islam]{mehedi2020systematic}
Syed Agha Hassnain Mohsan1~Md Mehedi, Alireza~Mazinani Hasan, Muhammad~Abubakar
  Sadiq, Hammad Akhtar, and Laraba Selsabil~Rokia Islam.
\newblock A systematic review on practical considerations, recent advances and
  research challenges in underwater optical wireless communication, 2020.

\bibitem[Cutler et~al.(2017)Cutler, Fowers, Kramer, and
  Peterson]{cutler2017dunking}
Ben Cutler, Spencer Fowers, Jeffrey Kramer, and Eric Peterson.
\newblock Dunking the data center.
\newblock \emph{IEEE Spectrum}, 54\penalty0 (3):\penalty0 26--31, 2017.

\bibitem[Simon(2018)]{simon2018project}
Kevin Simon.
\newblock Project natick-microsoft's self-sufficient underwater datacenters.
\newblock \emph{IndraStra Global}, 4\penalty0 (6):\penalty0 4, 2018.

\bibitem[Palitharathna et~al.(2020)Palitharathna, Suraweera, Godaliyadda,
  Herath, and Thompson]{palitharathna2020multi}
Kapila~WS Palitharathna, Himal~A Suraweera, Roshan~I Godaliyadda, Vijitha~R
  Herath, and John~S Thompson.
\newblock Multi-{AUV} placement for coverage maximization in underwater optical
  wireless sensor networks.
\newblock In \emph{Global Oceans 2020: Singapore--US Gulf Coast}, pages 1--8.
  IEEE, 2020.

\bibitem[Kaushal and Kaddoum(2016)]{kaushal2016underwater}
Hemani Kaushal and Georges Kaddoum.
\newblock Underwater optical wireless communication.
\newblock \emph{IEEE access}, 4:\penalty0 1518--1547, 2016.

\bibitem[Zhang et~al.(2020)Zhang, N'Doye, Ballal, {Al-Naffouri}, Alouini, and
  {Laleg-Kirati}]{ZNBAAL:20}
D.~Zhang, I.~N'Doye, T.~Ballal, T.-Y. {Al-Naffouri}, M.-S. Alouini, and T.-M.
  {Laleg-Kirati}.
\newblock Localization and tracking control using hybrid acoustic-optical
  communication for autonomous underwater vehicles.
\newblock \emph{IEEE Internet of Things Journal}, 7\penalty0 (10):\penalty0
  10048--10060, 2020.

\bibitem[N'Doye et~al.(2021)N'Doye, Zhang, Alouini, and Laleg-Kirati]{NZAL:21}
I.~N'Doye, D.~Zhang, M.-S. Alouini, and T.-M. Laleg-Kirati.
\newblock Establishing and maintaining a reliable optical wireless
  communication in underwater environment.
\newblock \emph{{IEEE} Access}, 9:\penalty0 62519--62531, 2021.

\bibitem[Saeed et~al.(2018)Saeed, Celik, Alouini, and
  Al-Naffouri]{saeed2018performance}
Nasir Saeed, Abdulkadir Celik, Mohamed-Slim Alouini, and Tareq~Y Al-Naffouri.
\newblock Performance analysis of connectivity and localization in multi-hop
  underwater optical wireless sensor networks.
\newblock \emph{IEEE Transactions on Mobile Computing}, 18\penalty0
  (11):\penalty0 2604--2615, 2018.

\bibitem[Zedini et~al.(2019)Zedini, Oubei, Kammoun, Hamdi, Ooi, and
  Alouini]{zedini2019unified}
Emna Zedini, Hassan~Makine Oubei, Abla Kammoun, Mounir Hamdi, Boon~S Ooi, and
  Mohamed-Slim Alouini.
\newblock Unified statistical channel model for turbulence-induced fading in
  underwater wireless optical communication systems.
\newblock \emph{IEEE Transactions on Communications}, 67\penalty0 (4):\penalty0
  2893--2907, 2019.

\bibitem[N'Doye et~al.(2020)N'Doye, Zhang, Zemouche, Rajamani, and
  {Laleg-Kirati}]{NZZRL:20}
I.~N'Doye, D.~Zhang, A.~Zemouche, R.~Rajamani, and T.-M. {Laleg-Kirati}.
\newblock A switched-gain nonlinear observer for {LED} optical communication.
\newblock In \emph{21st IFAC World Congress}, Berlin, Germany, 2020.

\bibitem[Solanki et~al.(2018{\natexlab{a}})Solanki, Al-Rubaiai, and
  Tan]{solanki2018extended}
Pratap~Bhanu Solanki, Mohammed Al-Rubaiai, and Xiaobo Tan.
\newblock Extended kalman filter-based active alignment control for {LED}
  optical communication.
\newblock \emph{IEEE/ASME Transactions on Mechatronics}, 23\penalty0
  (4):\penalty0 1501--1511, 2018{\natexlab{a}}.

\bibitem[Semcheddine and Bouchareb(2019)]{semcheddinerobust}
Samia Semcheddine and Hanane Bouchareb.
\newblock Robust control and state estimation of a three-stage anaerobic
  digestion process.
\newblock In \emph{Ecological Engineering and Environment Protection}, pages
  29--38, 2019.

\bibitem[Califano et~al.(2009)Califano, Monaco, and
  Normand-Cyrot]{califano2009canonical}
Claudia Califano, Salvatore Monaco, and Doroth{\'e}e Normand-Cyrot.
\newblock Canonical observer forms for multi-output systems up to coordinate
  and output transformations in discrete time.
\newblock \emph{Automatica}, 45\penalty0 (11):\penalty0 2483--2490, 2009.

\bibitem[Beineke et~al.(1997)Beineke, Sch{\"u}tte, and
  Grotstollen]{beineke1997comparison}
S~Beineke, F~Sch{\"u}tte, and H~Grotstollen.
\newblock Comparison of methods for state estimation and on-line identification
  in speed and position control loops.
\newblock In \emph{Proc. of the Intern. Conf. European Power Electronics},
  pages 3--364, 1997.

\bibitem[Kazantzis and Kravaris(2001)]{kazantzis2001discrete}
Nikolaos Kazantzis and Costas Kravaris.
\newblock Discrete-time nonlinear observer design using functional equations.
\newblock \emph{Systems \& Control Letters}, 42\penalty0 (2):\penalty0 81--94,
  2001.

\bibitem[Poulain et~al.(2008)Poulain, Praly, and Ortega]{poulain2008observer}
Fran{\c{c}}ois Poulain, Laurent Praly, and Romeo Ortega.
\newblock An observer for permanent magnet synchronous motors with currents and
  voltages as only measurements.
\newblock In \emph{2008 47th IEEE Conference on Decision and Control}, pages
  5390--5395. IEEE, 2008.

\bibitem[Fadlullah et~al.(2017)Fadlullah, Tang, Mao, Kato, Akashi, Inoue, and
  Mizutani]{fadlullah2017state}
Zubair~Md Fadlullah, Fengxiao Tang, Bomin Mao, Nei Kato, Osamu Akashi, Takeru
  Inoue, and Kimihiro Mizutani.
\newblock State-of-the-art deep learning: Evolving machine intelligence toward
  tomorrow’s intelligent network traffic control systems.
\newblock \emph{IEEE Communications Surveys \& Tutorials}, 19\penalty0
  (4):\penalty0 2432--2455, 2017.

\bibitem[Ramos et~al.(2020)Ramos, Di~Meglio, Morgenthaler, da~Silva, and
  Bernard]{ramos2020numerical}
Louise da~C Ramos, Florent Di~Meglio, Val{\'e}ry Morgenthaler, Lu{\'\i}s
  F~Figueira da~Silva, and Pauline Bernard.
\newblock Numerical design of luenberger observers for nonlinear systems.
\newblock In \emph{2020 59th IEEE Conference on Decision and Control (CDC)},
  pages 5435--5442. IEEE, 2020.

\bibitem[Peralez and Nadri(2021)]{Peralezl2021Deep}
Johan Peralez and Madiha Nadri.
\newblock Deep learning-based luenberger observers design for discrete-time
  nonlinear systems.
\newblock In \emph{2021 IEEE 60th Conference on Decision and Control (CDC)},
  2021.

\bibitem[Ghassemlooy et~al.(2012)Ghassemlooy, Popoola, and Rajbhandari]{GPR:12}
Z.~Ghassemlooy, W.~Popoola, and S.~Rajbhandari.
\newblock \emph{Optical Wireless Communications: System and Channel Modelling
  with {MATLAB}}.
\newblock CRC Press, Berlin, 1st edition, 2012.

\bibitem[Solanki et~al.(2018{\natexlab{b}})Solanki, {Al-Rubaiai}, and
  Tan]{SAT:18}
P.~B. Solanki, M.~{Al-Rubaiai}, and X.~Tan.
\newblock Extended {K}alman filter-based active alignment control for {LED}
  optical communication.
\newblock \emph{IEEE/ASME Transactions on Mechatronics}, 23\penalty0
  (4):\penalty0 1501--1511, 2018{\natexlab{b}}.

\bibitem[Bernard and Andrieu(2019)]{BeA:19}
P.~Bernard and V.~Andrieu.
\newblock Luenberger observers for nonautonomous nonlinear systems.
\newblock \emph{{IEEE} Trans. Automatic Control}, 64\penalty0 (1):\penalty0
  270--281, 2019.

\end{thebibliography}

\smallskip
\end{document}